\pdfoutput=1
\RequirePackage{fixltx2e}
\documentclass[aps,prl,reprint,floatfix,showpacs]{revtex4-1}

\usepackage{amsmath,amsfonts} %
\usepackage{microtype} %
\usepackage{wasysym} %
\usepackage{graphicx} %
\usepackage[dvipsnames,table]{xcolor} %
\usepackage[acronym]{glossaries} %
\usepackage{dcolumn} %
\usepackage{bm} %
\usepackage{ifpdf} %
\usepackage{braket} %
\usepackage{siunitx} %
\usepackage{fancyhdr} %
\usepackage{etoolbox} %
\usepackage{natbib} %
\usepackage[byname]{smartref} %
\usepackage[breaklinks, %
colorlinks, linkcolor=Blue, citecolor=Blue, urlcolor=Blue]{hyperref} %
\AtBeginDocument{\usepackage{booktabs}} 

\newcommand{\bfr}{\mathbf{r}}

\hypersetup{ %
  pdftitle={Reduction of Electronic Wavefunctions to Kohn--Sham
    Effective Potentials}, %
  pdfauthor={Ilya G. Ryabinkin, Sviataslau V. Kohut, and Viktor N.
    Staroverov}, %
  pdfsubject={Fundamentals in DFT}, %
  pdfkeywords={} %
}

\newacronym{DFT}{DFT}{density functional theory} %
\newacronym{CI}{CI}{configuration interaction} %
\newacronym{FCI}{FCI}{full configuration interaction} %
\newacronym{MCSCF}{MCSCF}{multiconfigurational self-consistent field}
\newacronym{CAS}{CAS}{complete active space} %
\newacronym{HOMO}{HOMO}{highest occupied molecular orbital} %
\newglossaryentry{RDM}{%
  type = \acronymtype, %
  name = {RDM}, %
  description = {reduced density matrix}, %
  text = {RDM}, %
  first = {reduced density matrix (RDM)}, %
  firstplural = {reduced density matrices (RDMs)}, %
  plural = {RDMs}} %

\begin{document}
\title{Reduction of Electronic Wavefunctions to Kohn--Sham Effective
  Potentials}

\author{Ilya G. Ryabinkin} %
\affiliation{Department of Physical and Environmental Sciences,
  University of Toronto Scarborough, Toronto, Ontario, M1C\,1A4,
  Canada}

\author{Sviataslau V. Kohut} %
\author{Viktor N. Staroverov} %
\email{vstarove@uwo.ca} %
\affiliation{Department of Chemistry, The University of Western
  Ontario, London, Ontario N6A\,5B7, Canada}

\date{\today}

\begin{abstract}
  A method for calculating the Kohn--Sham exchange-correlation
  potential, $v_\text{XC}(\bfr)$, from a given electronic wavefunction
  is devised and implemented. It requires on input one- and
  two-electron density matrices and involves construction of the
  generalized Fock matrix. The method is free from numerical
  limitations and basis-set artifacts of conventional schemes for
  constructing $v_\text{XC}(\bfr)$ in which the potential is recovered
  from a given electron density, and is simpler than various many-body
  techniques. The chief significance of this development is that it
  allows one to directly probe the functional derivative of the true
  exchange-correlation energy functional and to rigorously test and
  improve various density-functional approximations.
\end{abstract}

\glsresetall


\maketitle

The Kohn--Sham density-functional theory~\cite{Kohn:1965/pr/A1133} is
the most widely used method for electronic structure calculations of
molecules and solids. In this method, the ground-state energy of a
system is treated as a functional of the electron density $\rho(\bfr)$
and then partitioned in such a way that only one term, the
exchange-correlation energy $E_\text{XC}[\rho]$, remains unknown.
Application of the variational principle to the total energy
functional leads to a one-electron Schr\"{o}dinger equation with an
effective Hamiltonian that includes the system's electrostatic
potential and the exchange-correlation potential,
$v_\text{XC}([\rho];\bfr) = \delta
E_\text{XC}[\rho]/\delta\rho(\bfr)$. While the exact
$E_\text{XC}[\rho]$ can be written only in implicit
form~\cite{Levy:1979/pnas/6062}, its functional derivative
$v_\text{XC}([\rho];\bfr)$ can in principle be computed and visualized
as a \textit{function} of $\bfr$ for any particular non-interacting
$v$-representable density. High-quality Kohn--Sham potentials are used
for testing density-functional
approximations~\cite{Menconi:2001/jcp/3958}, accurate description of
electronic excitations~\cite{vanMeer:2014/jctc/4432}, and other
purposes.

Most existing methods for generating exact exchange-correlation
potentials fit the function $v_\text{XC}(\bfr)$ to a given
ground-state $\rho(\bfr)$ via the Kohn--Sham equations either by
iterative updates~\cite{Wang:1993/pra/R1591, vanLeeuwen:1994/pra/2421,
  Gritsenko:1995/pra/1870, Ryabinkin:2012/jcp/164113} or through some
constrained optimization~\cite{Zhao:1994/pra/2138,
  Tozer:1996/jcp/9200, Wu:2003/jcp/2498}. The target densities are
usually obtained from \textit{ab initio} wavefunctions which are
themselves discarded. Because small changes in $\rho(\bfr)$ can induce
large changes in $v_\text{XC}(\bfr)$~\cite{Savin:2003/IJQC/166},
potential-reconstruction methods that use only $\rho(\bfr)$ as input
suffer from numerical instabilities. Moreover, electron densities
generated using ubiquitous Gaussian basis sets correspond to
exchange-correlation potentials that wildly oscillate and
diverge~\cite{Schipper:1997/TCA/16, Mura:1997/JCP/9659,
  Gaiduk:2013/JCTC/3959, Kananenka:2013/JCP/074112}, a result that is
formally correct but unwanted. Kohn--Sham potentials can be also
constructed by many-body methods~\cite{Bartlett:2005/JCP/034104,
  Grabowski:2011/JCP/114111, Hesselmann:2005/JCP/244108,
  Jiang:2006/JCP/184108, Fabiano:2007/JCP/214102}, but these
techniques are quite elaborate and often require solving an integral
equation for $v_\text{XC}(\bfr)$, which is a challenge by itself.

Here, we propose a radically different method for computing the
exchange-correlation potential of a given many-electron system, which
avoids the above difficulties. In this method, the functional
derivative of the exact $E_\text{XC}[\rho]$ is obtained directly from
the system's electronic wavefunction. The approach represents a
nontrivial generalization of our technique for constructing Kohn--Sham
potentials corresponding to Hartree--Fock (HF) electron
densities~\cite{Ryabinkin:2013/PRL/013001, Kohut:2014/JCP/18A535} and
is conceptually related to the wavefunction-based analysis of
Kohn--Sham potentials developed by Baerends and
co-workers~\cite{Buijse:1989/PRA/4190, Gritsenko:1994/JCP/8955,
  Gritsenko:1996/JCP/8535, Gritsenko:1996/PRA/1957,
  Schipper:1998/PRA/1729}.

The basic idea of our approach is to derive two expressions for the
local electron energy balance, one of which originates from the
Kohn--Sham equations, the other from the Schr\"{o}dinger equation.
When one expression is subtracted from the other under the assumption
that the Kohn--Sham and wavefunction-based densities are equal, the
system's electrostatic potentials cancel out and the difference gives
an explicit formula for $v_\text{XC}(\bfr)$. For simplicity, the
treatment presented in this Letter is restricted to electronic singlet
ground states described with closed-shell Kohn--Sham determinants, and
assumes that all basis functions and orbitals are real (although the
notation for complex conjugate is retained).

Accomplishing the first part of this plan is easy. In the Kohn--Sham
scheme, the ground-state density of a singlet $N$-electron system is
obtained as $\rho^\text{KS}(\bfr) = \sum_{i} n_i|\phi_i(\bfr)|^2$,
where $n_i = 0$ or 2 are occupation numbers of the corresponding
Kohn--Sham orbitals ($N = \sum_i n_i$). The orbitals are obtained by
solving the equation
\begin{equation}
  \label{eq:KS}
  \left[ -\frac{1}{2}\nabla^2 + v(\bfr) + v_\text{H}^\text{KS}(\bfr)
    + v_\text{XC}(\bfr) \right]
  \phi_i(\bfr) = \epsilon_i \phi_i(\bfr),
\end{equation}
where $v(\bfr)$ is the electrostatic potential of the nuclei and
$v_\text{H}^\text{KS}(\bfr)$ is the electrostatic potential of
$\rho^\text{KS}(\bfr)$. If we multiply Eq.~\eqref{eq:KS} by
$n_i\phi_i^*(\bfr)$, sum over $i$, and divide through by
$\rho^\text{KS}(\bfr)$, we obtain
\begin{equation}
  \label{eq:v-KS}
  \frac{\tau_L^\text{KS}(\bfr)}{\rho^\text{KS}(\bfr)}
  + v(\bfr) + v_\text{H}^\text{KS}(\bfr) + v_\text{XC}(\bfr)
  = \bar{\epsilon}^\text{KS}(\bfr),
\end{equation}
where $\tau_L^\text{KS}(\bfr)=-(1/2) \sum_{i} n_i
\phi_i^{*}(\bfr)\nabla^2\phi_i(\bfr)$ is the Kohn--Sham kinetic energy
density and
\begin{equation}
  \label{eq:eps-KS}
  \bar{\epsilon}^\text{KS}(\bfr)
  = \frac{1}{\rho^\text{KS}(\bfr)}
  \sum_{i} n_i\epsilon_i |\phi_i(\bfr)|^2
\end{equation}
is the average local Kohn--Sham orbital
energy~\cite{Bulat:2009/JPCA/1384}.

The second part of the plan is to reduce the $N$-electron
Schr\"odinger equation to a local energy balance expression analogous
to Eq.~\eqref{eq:v-KS}. There is more than one way to do this. Holas
and March~\cite{Holas:1997/IJQC/263} had considered a contracted
Schr\"{o}dinger equation for this purpose, but their proposal led to a
complicated integral equation for $v_\text{XC}(\bfr)$ involving the
three-particle \gls{RDM}. The Baerends
group~\cite{Buijse:1989/PRA/4190, Gritsenko:1994/JCP/8955,
  Gritsenko:1996/JCP/8535, Gritsenko:1996/PRA/1957,
  Schipper:1998/PRA/1729} used an expression involving
$(N-1)$-electron conditional amplitudes. The method we propose here is
motivated by L\"{o}wdin's approach~\cite{Lowdin:1955/PR/1474} to the
problem of finding the optimal finite one-electron basis set for a
\gls{CI} expansion.

Suppose we have an $N$-electron ground-state wavefunction $\Psi$
expressed in terms of orthonormal orbitals $\{\psi_i\}$. Then the
total electronic energy may be written as
\begin{equation}
  \label{eq:E-RDM-matrix}
  E = \sum_{ij} \gamma_{ij} \langle \psi_j|\hat{h}|\psi_i\rangle
  + \sum_{ikjl} \Gamma_{ikjl} \langle
  \psi_j\psi_l|r_{12}^{-1}|\psi_i\psi_k\rangle,
\end{equation}
where $\hat{h}(\bfr) = -(1/2)\nabla^2 + v(\bfr)$ is the one-electron
core Hamiltonian, $\gamma_{ij} = \sum_\sigma
\langle\Psi|\hat{a}_{j\sigma}^\dagger \hat{a}_{i\sigma}|\Psi\rangle$
($\sigma = \alpha,\beta$ is the spin index) are matrix elements of the
spin-free one-particle \gls{RDM}, and $\Gamma_{ikjl} = (1/2)
\sum_{\sigma\sigma'} \langle\Psi|
\hat{a}_{j\sigma}^\dagger\hat{a}_{l\sigma'}^\dagger
\hat{a}_{k\sigma'}\hat{a}_{i\sigma}| \Psi\rangle$ are matrix elements
of the spin-free two-particle \gls{RDM}.

Our objective is to turn Eq.~\eqref{eq:E-RDM-matrix} into a local
energy balance equation. We start by minimizing $E$ with respect to
the functions $\{\psi_i\}$, subject to the constraint
$\langle\psi_j|\psi_i\rangle = \delta_{ji}$, while keeping
$\gamma_{ij}$ and $\Gamma_{ikjl}$ fixed. The corresponding
Euler--Lagrange equation is
\begin{equation}
  \label{eq:delta-E}
  \frac{\delta E}{\delta\psi_j^*(\bfr)}
  = \sum_{i} \lambda_{ij} \psi_i(\bfr),
\end{equation}
where $\lambda_{ij}$ are yet undetermined Lagrange multipliers. We
evaluate the functional derivative in Eq.~\eqref{eq:delta-E}, multiply
the result by $\psi_j^*(\bfr')$, sum over $j$, and obtain
\begin{equation}
  \label{eq:G-subst}
  \hat{h}(\bfr) \gamma(\bfr,\bfr')
  + 2 \int \frac{\Gamma(\bfr,\bfr_2;\bfr',\bfr_2)}{|\bfr-\bfr_2|} \, d\bfr_2
  = \sum_{ij} \lambda_{ij} \psi_i(\bfr) \psi_j^{*}(\bfr').
\end{equation}
where
\begin{equation}
  \gamma(\bfr,\bfr') = \sum_{ij} \gamma_{ij} \psi_i(\bfr) \psi_j^*(\bfr')
\end{equation}
and
\begin{equation}
  \Gamma(\bfr,\bfr_2;\bfr',\bfr_2') =
  \sum_{ikjl} \Gamma_{ikjl} \psi_i(\bfr) \psi_k(\bfr_2)
  \psi_j^{*}(\bfr') \psi_l^{*}(\bfr_2')
\end{equation}
are the coordinate representations of the spin-free one- and
two-particle \glspl{RDM}, respectively.

We denote the left-hand side of Eq.~\eqref{eq:G-subst} by
$G(\bfr,\bfr')$ and treat it as the kernel of an integral operator
defined by
\begin{equation}
  \label{eq:G-def}
  \hat{G}\psi_j(\bfr) = \int G(\bfr,\bfr')\psi_j(\bfr')\,d\bfr'.
\end{equation}
Then $\lambda_{ij}$ can be determined from Eqs.~\eqref{eq:G-subst} and
\eqref{eq:G-def} as
\begin{equation}
  \lambda_{ij} = \langle \psi_i|\hat{G}|\psi_j\rangle.
\end{equation}
The operator $\hat{G}$, known as the generalized Fock operator or
orbital Lagrangian, arises in various problems of quantum
chemistry~\cite{Lowdin:1955/PR/1474, Hinze:1973/JCP/6424,
  Day:1974/IJQC/501, Morrison:1992/JCC/1004, Book/Helgaker:2000}.

For our purposes, we need only the $\bfr = \bfr'$ part of
Eq.~\eqref{eq:G-subst} which after division by $\rho^\text{WF}(\bfr) =
\gamma(\bfr,\bfr)$ becomes
\begin{equation}
  \label{eq:G-diag}
  \frac{\tau_L^\text{WF}(\bfr)}{\rho^\text{WF}(\bfr)}
  + v(\bfr) + \frac{2}{\rho^\text{WF}(\bfr)}
  \int \frac{P(\bfr,\bfr_2)}{|\bfr-\bfr_2|}\, d\bfr_2
  = \bar{\epsilon}^\text{WF}(\bfr),
\end{equation}
where $\tau_L^\text{WF}(\bfr) = -(1/2) \left[ \nabla^2
  \gamma(\bfr,\bfr') \right]_{\bfr'=\bfr}$ is the interacting kinetic
energy density, $P(\bfr,\bfr_2) = \Gamma(\bfr,\bfr_2;\bfr,\bfr_2)$ is
the pair function, and
\begin{equation}
  \bar{\epsilon}^\text{WF}(\bfr) = \frac{1}{\rho^\text{WF}(\bfr)}
  \sum_{ij} \lambda_{ij}\psi_i(\bfr)\psi_j^*(\bfr).
  \label{eq:Grho}
\end{equation}
One can always write the pair function as
\begin{equation}
  \label{eq:P}
  P(\bfr,\bfr_2) = \frac{1}{2} \rho^\text{WF}(\bfr) \left[
    \rho^\text{WF}(\bfr_2) + \rho_\text{XC}^\text{WF}(\bfr,\bfr_2)
  \right],
\end{equation}
which defines $\rho_\text{XC}^\text{WF}(\bfr,\bfr_2)$, the
exchange-correlation hole density. Substituting Eq.~\eqref{eq:P} into
Eq.~\eqref{eq:G-diag} we obtain
\begin{equation}
  \label{eq:G-explicit}
  \frac{\tau_L^\text{WF}(\bfr)}{\rho^\text{WF}(\bfr)}
  + v(\bfr) + v_\text{H}^\text{WF}(\bfr) + v_\text{S}^\text{WF}(\bfr)
  = \bar{\epsilon}^\text{WF}(\bfr),
\end{equation}
where $v_\text{H}^\text{WF}(\bfr)$ is the electrostatic potential of
$\rho^\text{WF}(\bfr)$ and
\begin{equation}
  \label{eq:vXC-S}
  v_\text{S}^\text{WF}(\bfr) = \int \frac{\rho_\text{XC}^\text{WF}(\bfr,\bfr_2)}
  {|\bfr-\bfr_2|}\,d\bfr_2
\end{equation}
is the Slater exchange-correlation-charge
potential~\cite{Slater:1953/PR/528}. Equation \eqref{eq:G-explicit} is
the wavefunction counterpart of Eq.~\eqref{eq:v-KS}.

Observe that the sum in Eq.~\eqref{eq:Grho} does not change if we
replace every $\lambda_{ij}$ with $\lambda_{ji}^*$. This means that
$\bar{\epsilon}^\text{WF}(\bfr)$ is determined by the Hermitian
(symmetric) part of $\hat{G}$. If desired, one can define the
self-adjoint operator $\hat{F} = (\hat{G}+\hat{G}^\dagger)/2$ and
solve the Hermitian eigenvalue problem $\hat{F}f_i(\bfr) =
\lambda_if_i(\bfr)$. This optional step allows one to cast
Eq.~\eqref{eq:Grho} as
\begin{equation}
  \label{eq:eps-WF}
  \bar{\epsilon}^\text{WF}(\bfr)
  = \frac{1}{\rho^\text{WF}(\bfr)}
  \sum_{i} \lambda_{i} |f_i(\bfr)|^2,
\end{equation}
which is formally analogous to Eq.~\eqref{eq:eps-KS}. The quantity
$\bar{\epsilon}^\text{WF}(\bfr)$ as given by Eq.~\eqref{eq:eps-WF} was
introduced by us earlier under the name of ``average local electron
energy"~\cite{Ryabinkin:2014/JCP/084107}.

Now let us subtract Eq.~\eqref{eq:G-explicit} from
Eq.~\eqref{eq:v-KS}, substitute the identity $\tau_L =
\tau-\nabla^2\rho/4$ for $\tau_L^\text{KS}$ and for $\tau_L^\text{WF}$
with $\tau^\text{KS} = (1/2)\sum_{i} n_i|\nabla\phi_i|^2$ and
$\tau^\text{WF}(\bfr) = (1/2)\left[\nabla_{\bfr'}\nabla_{\bfr}
  \gamma(\bfr,\bfr')\right]_{\bfr'=\bfr}$, and apply the condition
$\rho^\text{KS}(\bfr) = \rho^\text{WF}(\bfr)$. This yields the central
equation of this work:
\begin{equation}
  \label{eq:vXC}
  v_\text{XC}(\bfr) = v_\text{S}^\text{WF}(\bfr)
  + \bar{\epsilon}^\text{KS}(\bfr) - \bar{\epsilon}^\text{WF}(\bfr)
  + \frac{\tau^\text{WF}(\bfr)}{\rho^\text{WF}(\bfr)}
  - \frac{\tau^\text{KS}(\bfr)}{\rho^\text{KS}(\bfr)}.
\end{equation}
Since $\tau^\text{KS}$ and $\bar{\epsilon}^\text{KS}$ are initially
unknown, Eq.~\eqref{eq:vXC} must be solved iteratively in conjunction
with the Kohn--Sham equations. The transition from $\tau_L$ to $\tau$
is not strictly necessary but beneficial for numerical calculations
because $\tau$ does not diverge at the nuclei as does $\tau_L$.

Note that as $r\to\infty$, the term $v_\text{S}^\text{WF}$ vanishes,
but the other ingredients remain nonzero: $\bar{\epsilon}^\text{KS}$,
$\tau_L^\text{KS}/\rho^\text{KS}$, and
$-\tau^\text{KS}/\rho^\text{KS}$ approach
$\epsilon_\text{HOMO}$~\cite{Ayers:2002/IJQC/309}, while
$\bar{\epsilon}^\text{WF}$, $\tau_L^\text{WF}/\rho^\text{WF}$, and
$-\tau^\text{WF}/\rho^\text{WF}$ approach
$-I_\text{min}$~\cite{Ryabinkin:2014/JCP/084107}, where $I_\text{min}$
is the first ionization energy of the system as determined by the
extended Koopmans theorem~\cite{Morrell:1975/JCP/549}. To ensure that
$v_\text{XC}(\bfr)$ as given by Eq.~\eqref{eq:vXC} properly vanishes
at infinity, we shift all current values of $\epsilon_i$ in each
Kohn--Sham iteration to satisfy the condition
\begin{equation}
  \epsilon_\text{HOMO} = -I_\text{min},
\end{equation}
which also imparts $\rho^\text{KS}(\bfr)$ with proper asymptotic
decay.

The proposed algorithm is as follows.
\begin{enumerate}
  \setlength{\itemsep}{0.5ex} %
  \setlength{\parskip}{0ex} %
  \setlength{\parsep}{0ex} %

\item Obtain a wavefunction for the system of interest. Calculate
  $\rho^\text{WF}$, $\tau^\text{WF}$, $v_\text{S}^\text{WF}$,
  $\bar{\epsilon}^\text{WF}$, and $I_\text{min}$.

\item Generate an initial guess for the occupied Kohn--Sham orbitals
  $\{\phi_i\}$ and their eigenvalues $\{\epsilon_i\}$.

\item Using the current guess for $\{\phi_i\}$ and shifted
  $\{\epsilon_i\}$, construct the potential $v_\text{XC}$ by
  Eq.~\eqref{eq:vXC}.

\item Solve the Kohn--Sham equations using the current $v_\text{XC}$
  and the same basis as in step 1. This gives new sets $\{\phi_i\}$
  and $\{\epsilon_i\}$.

\item Return to step 3 and iterate until the potential $v_\text{XC}$
  is self-consistent.

\end{enumerate}

The method was implemented in the \textsc{gaussian 09} suite of
programs~\cite{G09-short}, which already contains subroutines for
constructing the generalized Fock matrix as part of the \gls{MCSCF}
module. The values of $I_\text{min}$ were computed as in
Ref.~\citenum{Morrison:1992/JCC/1004}, while $\rho^\text{WF}$ and
$\tau^\text{WF}$ were assembled from natural orbitals. Any reasonable
density-functional approximation may be used to generate an initial
guess for $\{\phi_i\}$ and $\{\epsilon_i\}$. The potential was
considered converged when all Kohn--Sham density matrix elements from
consecutive iterations differed by less than $10^{-10}$ in the
root-mean-square sense. The method works best with basis sets that are
not heavily contracted in the core region.

\begin{figure}
  \centering
  \includegraphics[width=1.0\columnwidth]{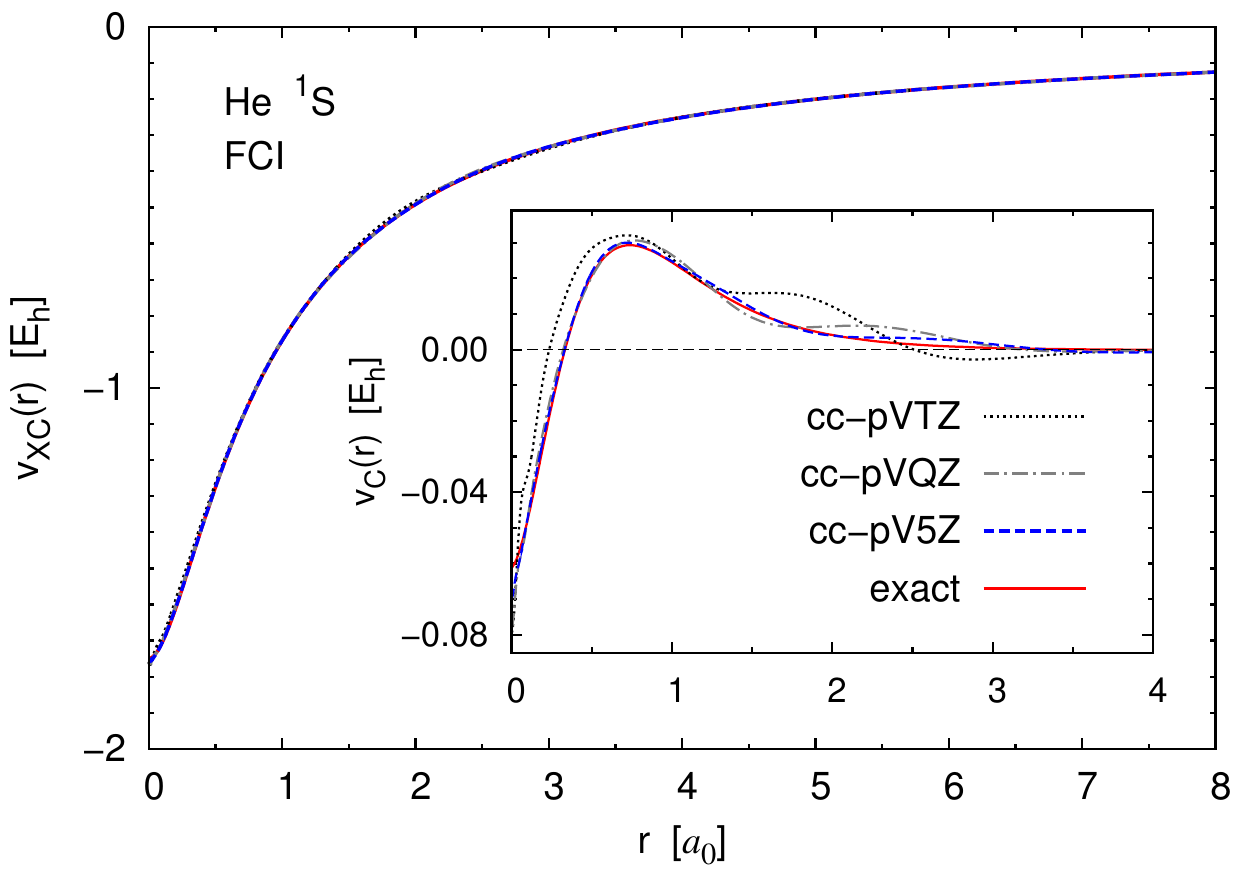}
  \caption{Exchange-correlation and correlation (inset) potentials for
    the He atom calculated from \protect\gls{FCI} wavefunctions using
    various basis sets.}
  \label{fig:1}
\end{figure}

\begin{figure}
  \centering
  \includegraphics[width=1.0\columnwidth]{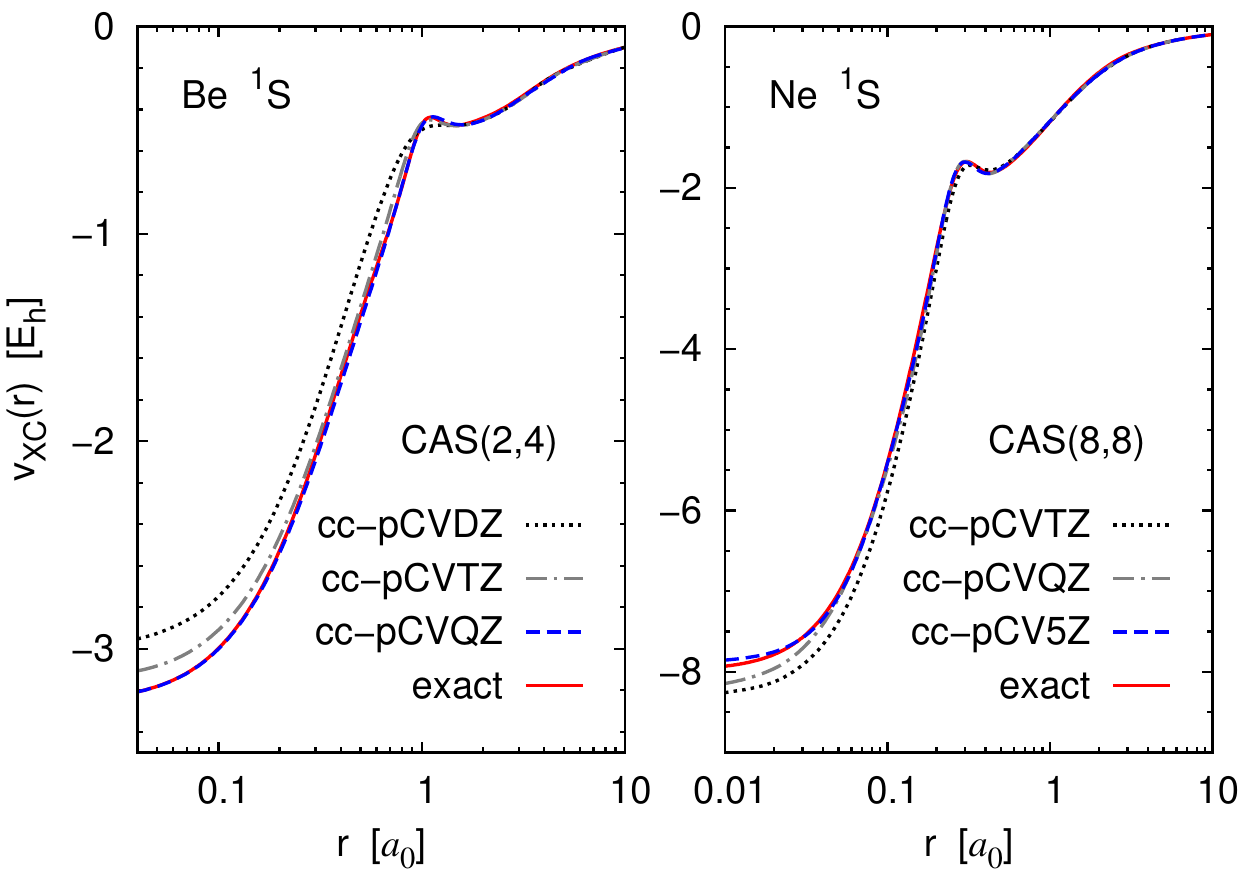}
  \caption{Exchange-correlation potentials for the Ne and Be atoms
    calculated from compact CASSCF wavefunctions using various basis
    sets.}
  \label{fig:2}
\end{figure}

\begin{table*}
  \caption{Characteristics of selected wavefunctions and the corresponding
    Kohn--Sham effective potentials (in atomic units).}
  \centering
  \begin{tabular*}{1.0\textwidth}{@{\extracolsep{\fill}}ll
      *{6}{S[detect-weight]}@{}} %
    \toprule
    System & Wavefunction
    & \multicolumn{1}{c}{$E_\text{tot}$}
    & \multicolumn{1}{c}{$I_\text{min}$}
    & \multicolumn{1}{c}{$T_s$}
    & \multicolumn{1}{c}{$T_c=T-T_s$}
    & \multicolumn{1}{c}{$E_\text{XC}^\text{KS}$}
    & \multicolumn{1}{c}{\quad $\Delta\rho$} \\
    \midrule
    He & FCI/cc-pVTZ        & -2.900232 & 0.9013 &  2.8571 & 0.0435 & -1.0550 & 0.00251 \\
    & FCI/cc-pVQZ           & -2.902411 & 0.9026 &  2.8652 & 0.0370 & -1.0645 & 0.00065 \\
    & FCI/cc-pV5Z           & -2.903152 & 0.9032 &  2.8668 & 0.0364 & -1.0662 & 0.00013 \\
    & Exact\footnotemark[1] & -2.903724 & 0.9037 &  2.8671 & 0.0366 & -1.0667 & \\
    Be & CAS(2,4)/cc-pCVDZ  & -14.61545 & 0.3485 & 14.4901 & 0.1333 & -2.6146 & 0.01729 \\
    & CAS(2,4)/cc-pCVTZ     & -14.61653 & 0.3489 & 14.5538 & 0.0619 & -2.6866 & 0.00493 \\
    & CAS(2,4)/cc-pCVQZ     & -14.61677 & 0.3490 & 14.5910 & 0.0258 & -2.7232 & 0.00547 \\
    & FCI/u-cc-pCVTZ        & -14.66370 & 0.3421 & 14.5956 & 0.0654 & -2.7715 & 0.00215 \\
    & Exact\footnotemark[1] & -14.66736 & 0.3426 & 14.5942 & 0.0732 & -2.7701 & \\
    \bottomrule
  \end{tabular*}
  \footnotetext[1]{Accurate estimates from Ref.~\citenum{Huang:1997/PRA/290} (He)
    and Ref.~\citenum{Filippi:1996/BOOK/295} (Be).}
  \label{tab:1}
\end{table*}

An added benefit of generating $v_\text{XC}(\bfr)$ from a wavefunction
is that one can readily obtain the corresponding exchange-correlation
energy, $E_\text{XC}^\text{KS}$, which is inaccessible when only the
electron density is known. We computed this energy as
$E_\text{XC}^\text{KS} = E_\text{XC}^\text{WF} + T_c$, where
$E_\text{XC}^\text{WF}$ is the \textit{ab initio} exchange-correlation
energy defined as $E_\text{XC}^\text{WF}=(1/2)\int
\rho^\text{WF}(\bfr) v_\text{S}^\text{WF}(\bfr)\,d\bfr$ and $T_c=T -
T_s$ is the difference between the \textit{ab initio} and Kohn--Sham
total kinetic energies, evaluated analytically. Also of interest is
the integrated density difference, $\Delta\rho = \int
|\rho^\text{KS}(\bfr)-\rho^\text{WF}(\bfr)|\,d\bfr$, evaluated for the
self-consistent $v_\text{XC}(\bfr)$. Because the condition
$\rho^\text{KS}(\bfr) = \rho^\text{WF}(\bfr)$ is imposed in our
approach only in the derivation of Eq.~\eqref{eq:vXC}, $\Delta\rho$
strictly vanishes only in the basis-set limit. Insistence on
reproducing $\rho^\text{WF}(\bfr)$ \textit{exactly} in Gaussian basis
sets would be misplaced because (i)~it brings out unwanted
oscillations and divergences of $v_\text{XC}(\bfr)$ and (ii)~the
potential that yields a given density in a finite basis is not unique
anyway~\cite{Harriman:1983/PRA/632,Staroverov:2006/JCP/141103}.

To test the method, we computed exchange-correlation potentials for
the three atoms (He, Be, and Ne) for which exact potentials are
available in the literature~\cite{Huang:1997/PRA/290,
  Filippi:1996/BOOK/295} using \acrfull{FCI} and \gls{CAS} SCF
wavefunctions and standard Gaussian basis sets~\cite{emsl-2}. For He,
already the potential extracted from the FCI wavefunction in the
cc-pVTZ basis set is very close to the exact $v_\text{XC}(\bfr)$, and
the cc-pVQZ and cc-pV5Z FCI exchange-correlation potentials are
visually indistinguishable from the benchmark (Fig.~\ref{fig:1} and
Table~\ref{tab:1}). Even the correlation potential for He,
$v_\text{C}(\bfr)=v_\text{XC}(\bfr)-v_\text{H}(\bfr)/2$, which is
almost two orders of magnitude smaller than $v_\text{XC}(\bfr)$, is
very accurate at the FCI/cc-pV5Z level (Fig.~\ref{fig:1}). For Be, the
sequence of potentials from CAS(2,4) wavefunctions quickly approaches
the exact $v_\text{XC}(\bfr)$ with increasing basis set size
(Fig.~\ref{fig:2}), as do the corresponding $T_s$ values
(Table~\ref{tab:1}). By contrast, $T_c$ and $E_\text{XC}^\text{KS}$
converge slowly because they depend not only on $v_\text{XC}(\bfr)$
but also on the accuracy of the wavefunction through the value of $T$.
Potentials for the Ne atom constructed from CAS(8,8) wavefunctions
also improve rapidly with the size of the basis set
(Fig.~\ref{fig:2}). Thus, even compact correlated wavefunctions can
produce accurate Kohn--Sham potentials, provided that the basis set is
of good quality.

\begin{figure}
  \centering
  \includegraphics[width=1.0\columnwidth]{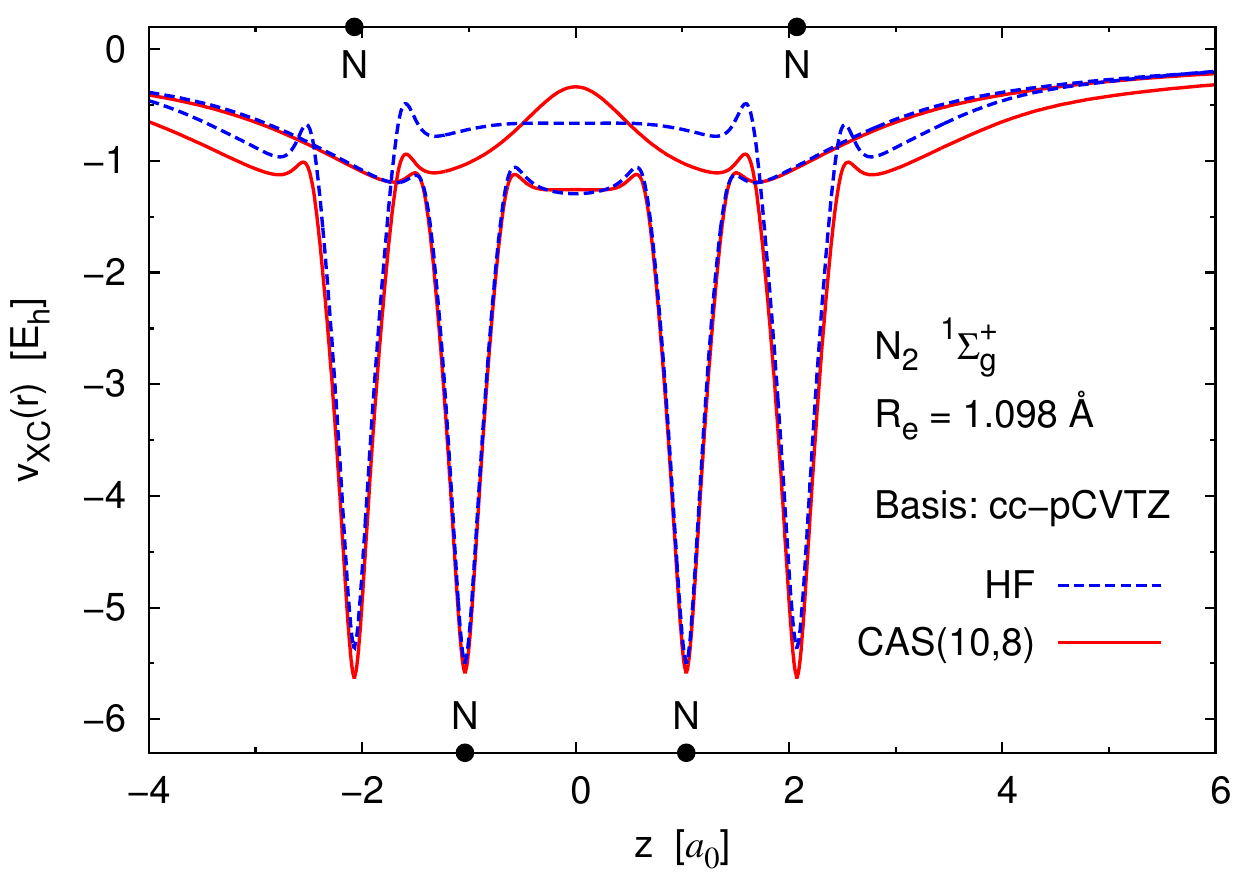}
  \caption{Exchange-correlation potentials for the N$_2$ molecule
    obtained from HF and valence CASSCF wavefunctions at the
    experimental equilibrium bond length and at $2R_e$.}
  \label{fig:3}
\end{figure}

\begin{figure}[!h]
  \centering
  \includegraphics[width=1.0\columnwidth]{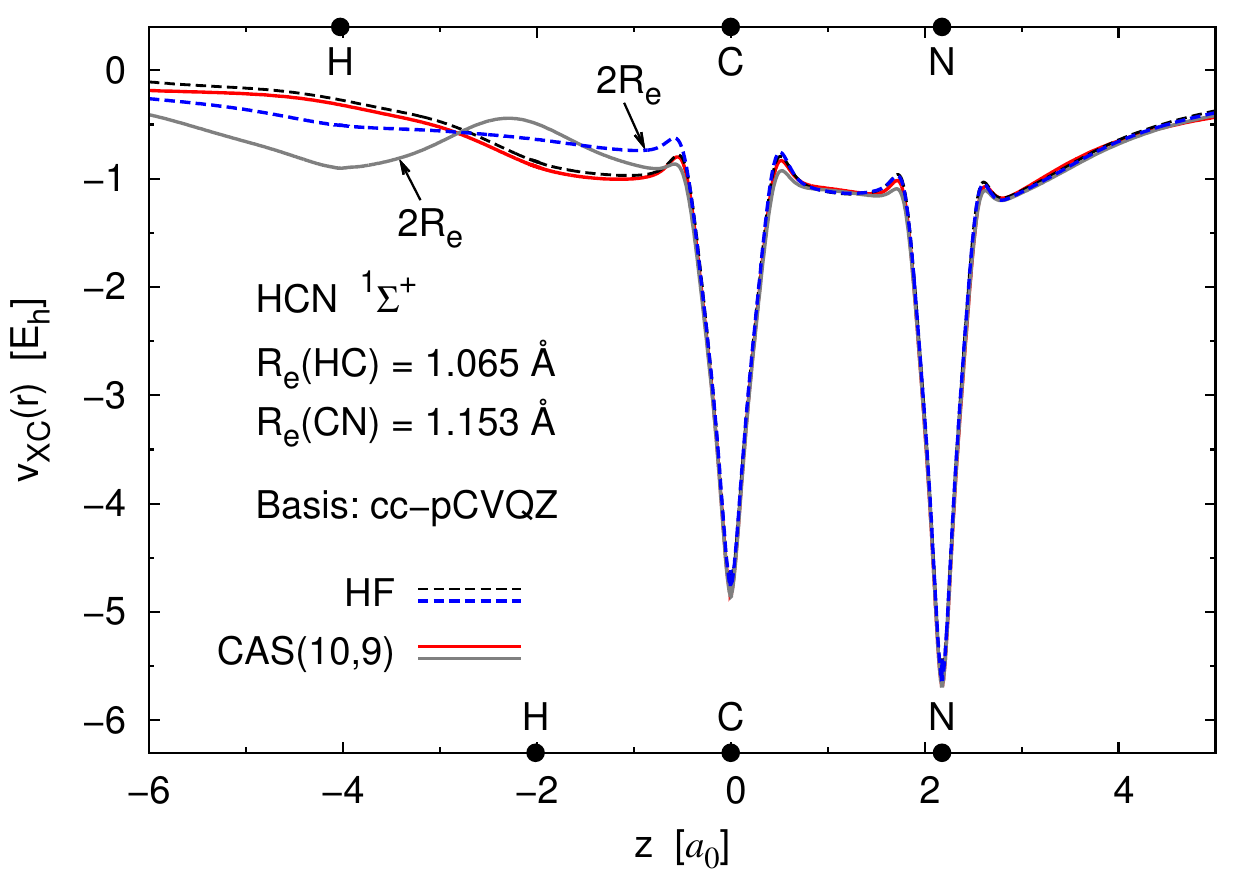}
  \caption{Exchange-correlation potentials for HCN obtained from HF
    and valence CASSCF wavefunctions at the experimental equilibrium
    geometry and with $R(\mbox{HC})=2R_e(\mbox{HC})$.}
  \label{fig:4}
\end{figure}

The method works equally well for molecules. It is known that, in
molecules, the onset of strong correlation induced by bond stretching
manifests itself in characteristic mid-bond peaks of
$v_\text{XC}(\bfr)$~\cite{Gritsenko:1996/PRA/1957,
  Gritsenko:1997/TCA/44, Gritsenko:1998/PRA/3450,
  Tempel:2009/JCTC/770}. Using our method, we readily reproduced these
peaks in a number of stretched diatomics exemplified by N$_2$
(Fig.~\ref{fig:3}). Exchange-correlation potentials for polyatomic
molecules can also be generated by our method (Fig.~\ref{fig:4}).

It is remarkable that Kohn--Sham potentials computed from
wavefunctions are always well-defined and free from spurious features.
Conventional methods for extracting $v_\text{XC}(\bfr)$ from
densities, when implemented in matrix form, would not deliver such
unambiguous results because there is no one-to-one correspondence
between densities and potentials in finite basis
sets~\cite{Harriman:1983/PRA/632}. Furthermore, when
density-to-potential mapping techniques are rigorously applied to
electron densities generated in Gaussian basis sets, one obtains
unphysical potentials~\cite{Schipper:1997/TCA/16, Mura:1997/JCP/9659,
  Gaiduk:2013/JCTC/3959, Kananenka:2013/JCP/074112}. Neither of these
complications affects our approach.

In conclusion, we have developed a practical method for folding a
many-electron wavefunction into the corresponding exchange-correlation
potential. The key ingredient of our approach is the generalized Fock
matrix which is commonly available in \textit{ab initio} codes as a
by-product of computing MCSCF wavefunctions, nuclear gradients, and
first-order properties. The method possesses several advantages over
existing techniques for constructing exchange-correlation potentials:
it delivers $v_\text{XC}(\bfr)$ in a simple analytic form, avoids the
ambiguity of associating a given electron density with a Kohn--Sham
potential in a finite basis set, is stable with respect to changes in
basis sets, convergence thresholds and other details of the
calculation, and produces potentials without oscillations and
divergences when using Gaussian basis sets. Further exploration of the
capabilities of our approach is under way.

\begin{acknowledgments}
  The authors thank Michael Frisch for help with the \textsc{gaussian}
  code and Cyrus Umrigar for providing the exchange-correlation
  potential benchmarks. This work was funded by the Natural Sciences
  and Engineering Research Council of Canada (NSERC) through the
  Discovery Grants Program. S.V.K.~acknowledges support from the
  Ontario Trillium Scholarship Program.
\end{acknowledgments}

\end{document}